\documentclass[twocolumn]{aa}  
\usepackage{graphicx}
\usepackage{color}
\usepackage{epsfig}
\usepackage{supertabular}
\def\etal{{\it et~al.}~}
\usepackage{txfonts}
\begin{document}
   \title{On the efficiency of field star capture by star clusters}


   \author{S. Mieske
          \inst{1}
          \and
          H. Baumgardt\inst{2}
          }

   \offprints{S. Mieske}

   \institute{European Southern Observatory, Karl-Schwarzschild-Strasse 2, 85748 Garching bei M\"unchen, Germany\\
              \email{smieske@eso.org}
         \and
     Argelander-Institut f\"ur Astronomie, Auf dem H\"ugel 71, 53121 Bonn, Germany\\\email{holger@astro.uni-bonn.de}}

   \date{}

 
  \abstract
   {An exciting recent finding regarding scaling relations among globular
clusters is the so-called 'blue tilt': clusters of
the blue sub-population follow a trend of redder colour with increasing luminosity.} 
   {In this paper we evaluate to which extent field star capture over a Hubble time changes the photometric properties of star clusters. Given that field stars
in early type giant galaxies are very metal-rich, their capture will make blue GCs redder and may in principle explain the 'blue tilt'.}
{We perform collisional N-body simulations to quantify the amount of
  field star capture occuring over a Hubble time to star clusters with
  10$^3$ to 10$^6$ stars. In the simulations we follow the orbits of
  field stars passing through a star cluster and calculate the energy
  change that the field stars experience due to gravitational
  interaction with cluster stars during one passage through the
  cluster. The capture condition is that their total energy after the
  passage is smaller than the gravitational potential at the cluster's
  tidal radius. By folding this with the fly-by rates of field stars
  with an assumed space density as in the solar neighbourhood and a
  range of velocity dispersions $\sigma$ (15 to 485 km$s^{-1}$), we
  derive estimates on the mass fraction of captured field stars as a
  function of environment.}
{We find that integrated over a Hubble time, the ratio between
  captured field stars and total number of clusters stars is very low
  ($\lesssim 10^{-4}$), even for the smallest field star velocity dispersion
  $\sigma=$15 km$s^{-1}$. This holds for star clusters in the
  mass range of both open clusters and globular clusters. We furthermore
  show that tidal friction has a negligible effect on the energy
  distribution of field stars after interaction with the cluster. We
  note that field star capture at the time of cluster formation, when
  the cluster potential increases with time, is more efficient. However, it
  cannot explain the trend that more massive star clusters are
  redder.}
   {Field star capture is not a probable mechanism
     for creating the colour-magnitude trend of metal-poor
     globular clusters.}


   \maketitle 
%

\section{Introduction}
An exciting recent finding regarding scaling relations among globular
clusters is that the colours of individual globular clusters (GCs) of
the blue sub-population are correlated with their luminosities. This
correlation is such that brighter globulars are redder (Harris et
al.~\cite{Harris06}, Mieske {\it et al.}~\cite{Mieske06}, Strader et
al.~\cite{Strade06}, Spitler {\it et al.}~\cite{Spitle06}, Cantiello et
al.~\cite{Cantie07}).  The amplitude of this 'blue tilt'  is about 0.03 to 0.04
mag in colour per mag in luminosity. Assuming coeval GCs, the trend implies a
relation between mass of the GCs and the luminosity weighted mean
metallicity of its member stars.  Various mechanisms have been discussed that may offer ways towards explain the trend, like self-enrichment (Strader et
al.~\cite{Spitle06}) or ``sample contamination'' by stripped nuclei of
dwarf galaxies (Harris {\it et al.}~\cite{Harris06}, Bekki et
al.~\cite{Bekki07}).

Self-enrichment, especially if pressure-induced (Mieske et
al.~\cite{Mieske06}, Parmentier~\cite{Parmen04}), may offer a
plausible way toward explaining the trend. Numerous authors have
discussed the possibility of star cluster self enrichment, but there
is a wide range of conclusions as to whether and to which extent it is
possible in GCs (e.g., Frank \& Gisler~\cite{Frank76},
Smith~\cite{Smith96}, Gnedin \etal~\cite{Gnedin02}, Parmentier \&
Gilmore~\cite{Parmen01}, Dopita \& Smith~\cite{Dopita86}, Morgan \&
Lake~\cite{Morgan89}, Thoul \etal~\cite{Thoul02}, Recchi \&
Danziger~\cite{Recchi05}, Prantzos \& Charbonnel~\cite{Prantz06}).

The presence of contaminators like stripped nuclei can probably not
give a satisfactory explanation for the trend. Given the Gaussian
shape of the blue colour peak (Bekki {\it et al.}~\cite{Bekki07}, Peng et
al.~\cite{Peng06}) over the observed magnitude range, one would
require the ``contaminators'' to actually dominate the GC sample. This
is unlikely, given the much smaller number of stripped nuclei expected
over a Hubble time (Bekki {\it et al.}~\cite{Bekki03}, Mieske {\it et
al.}~\cite{Mieske06}).

In Mieske {\it et al.}~(\cite{Mieske06}, M06 in the following) we indicate
that also the capture of field stars in giant elliptical galaxies can
in principle cause such colour-magnitude trends, including the
dependence of the trend on field star density that is detected in M06.
This is because the field star population is generally much redder
than the blue globular clusters. Assuming a non-linear dependence of
the capture efficiency on globular cluster mass, a colour-mass trend
will occur. The question is whether clusters can obtain a
sufficiently large population of captured field stars (several
percent) to create a notable trend. More generally, it is also of
interest to which extent field star capture may explain the multiple
populations detected in massive Milky Way globular clusters (e.g. Bedin et al.~\cite{Bedin04}, D'Antona et
al.~\cite{Danton05}, Piotto et al.~\cite{Piotto07}).

In this paper, we present collisional N-body simulations to quantify
the amount of field star capture occuring over a Hubble time to star
clusters with a time-invariant gravitational potential. The setup of
the simulations is presented in Sect.~\ref{setup}, the results are
shown in Sect.~\ref{results}.  The discussion in Sect.~\ref{discussion}
includes a comparison between our results and estimates on field star capture
in bulge globular clusters from Bica {\it et al.}~(\cite{Bica97}), which
are markedly different.

\label{Intro}
\section{Setup of simulations}
\label{setup}
We simulate the passage of field stars through a star cluster by means
of direct $N$-body simulations, using a fourth order Hermite scheme
with individual time-steps to follow the orbits of field and cluster
stars. For the sake of simplicity we do not include the tidal field of
the host galaxy in the calculations. This is justified since
interactions are restricted to the central, high density parts of the
cluster, where the host galaxy potential does not play a role. The
host galaxy properties are folded in later, when capture probabilities
are derived as a function of the ratio between tidal radius r$_{tid}$
and cluster half-mass radius r$_h$. We assume a time-invariant
gravitational potential for the star clusters, since the typical
crossing times of a few Myr are very small compared to mass loss time
scales (e.g. Baumgardt \& Makino~\cite{BM03}, Lamers et
al.~\cite{Lamers05}).
In our calculations, the interactions between field stars and cluster
stars are calculated at each time step, while the cluster stars do not
interact among themselves, but rather feel a smooth cluster potential.
 Allowing for direct interactions between the cluster stars would
  have increased the required computation time to prohibitively large
  values.  We have therefore not included this case in the present
  study.  It is in any case unlikely that direct interactions between
  cluster stars influence our results: the interaction of the field
  stars with the cluster stars happens on a crossing time, while
  interactions between cluster stars lead to orbital changes only on a
  relaxation time, which is at least a factor of 100 longer
  for the considered clusters.

We simulate three star clusters with particle numbers of N=$10^3$,
$10^4$, and 10$^5$, assuming a Plummer profile for the stellar density
distribution.  Three different values of the initial radial velocity
at infinity of the field stars v$_{\rm ini}$ are simulated, namely
v$_{\rm ini}$=0.1, 0.33, and 1 $\sigma_{\rm cluster}$, where
$\sigma_{\rm cluster}$ is the velocity dispersion of the star cluster.
For each particle number and field star velocity, we simulate the
passage of 10$^4$ field stars.

\begin{figure}[h!]
\begin{center}
  \epsfig{figure=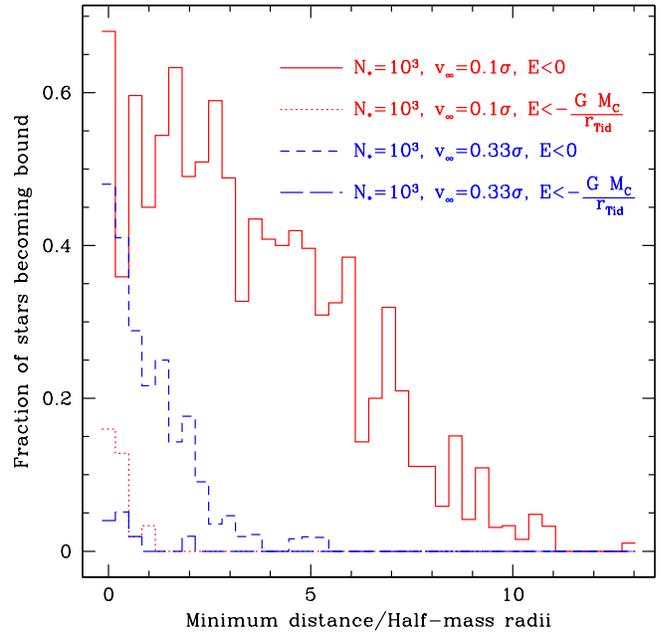,width=8.6cm}
\caption{The fraction of captured field stars vs. the minimum distance to the cluster centre for 
  different initial stellar velocities and capture criteria. For the
  dotted and long-dashed curves, a tidal radius of $r_{tid} = 20
  r_h$ was assumed, while the solid and short-dashed curves
  assume no tidal field of the galaxy.  For star clusters in tidal
  fields, significant capture of field stars occurs only for impact
  parameters $p$ smaller than 2$r_h$.}
\label{radius}
\end{center}
\end{figure}

The distance of closest approach $p(b)$ as a function of the initial
impact parameter $b$ at infinity is given by
\begin{equation}
p(b)= - G*M_c/v_{\rm ini}^2 + \sqrt{(G*M_c/v_{\rm ini}^2)^2 + b^2} 
\end{equation}
where $M_c$ is the mass of the cluster and $G$ the gravitational
constant.  For the direct numerical calculation of a given field
star's trajectory, we restrict the distance of closest passage $p(b)$
to smaller than two half-mass radii of the star cluster since larger
impact parameters do not result in any notable interaction between
cluster and field stars and can hence be excluded from our runs (see
Fig.~\ref{radius}).  The initial impact parameter $b$ of each star is
chosen randomly within the surface area that corresponds to
$p<2*r_{\rm h}$. The direct numerical calculation for a given field
star is started at a distance of 2 $r_{\rm tid}$. The velocity and
position at this distance are calculated by analytically integrating
the orbit, assuming a Keplerian ellipse with parameters based on $b$
and v$_{\rm ini}$. After following each individual passage, a star is
considered as captured if its total energy $E$ after the interaction
is smaller than the potential energy at the tidal radius: $E<\frac{-G
  M_c}{r_{\rm tid}}$.


\section{Results}
\label{results}
\subsection{Capture probabilities}
\begin{figure*}[h!]
\begin{center}
  \epsfig{figure=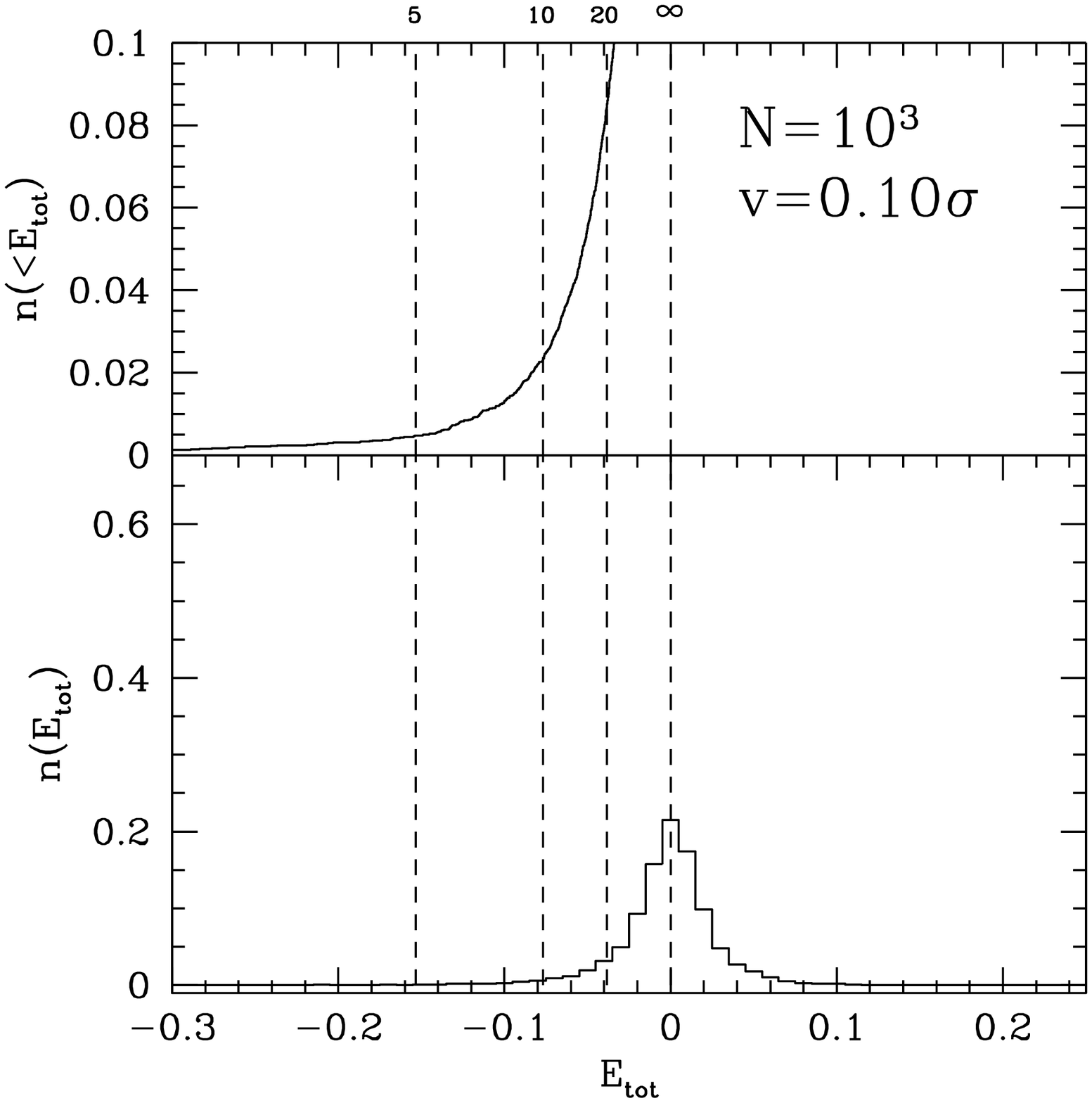,width=6.8cm}
\epsfig{figure=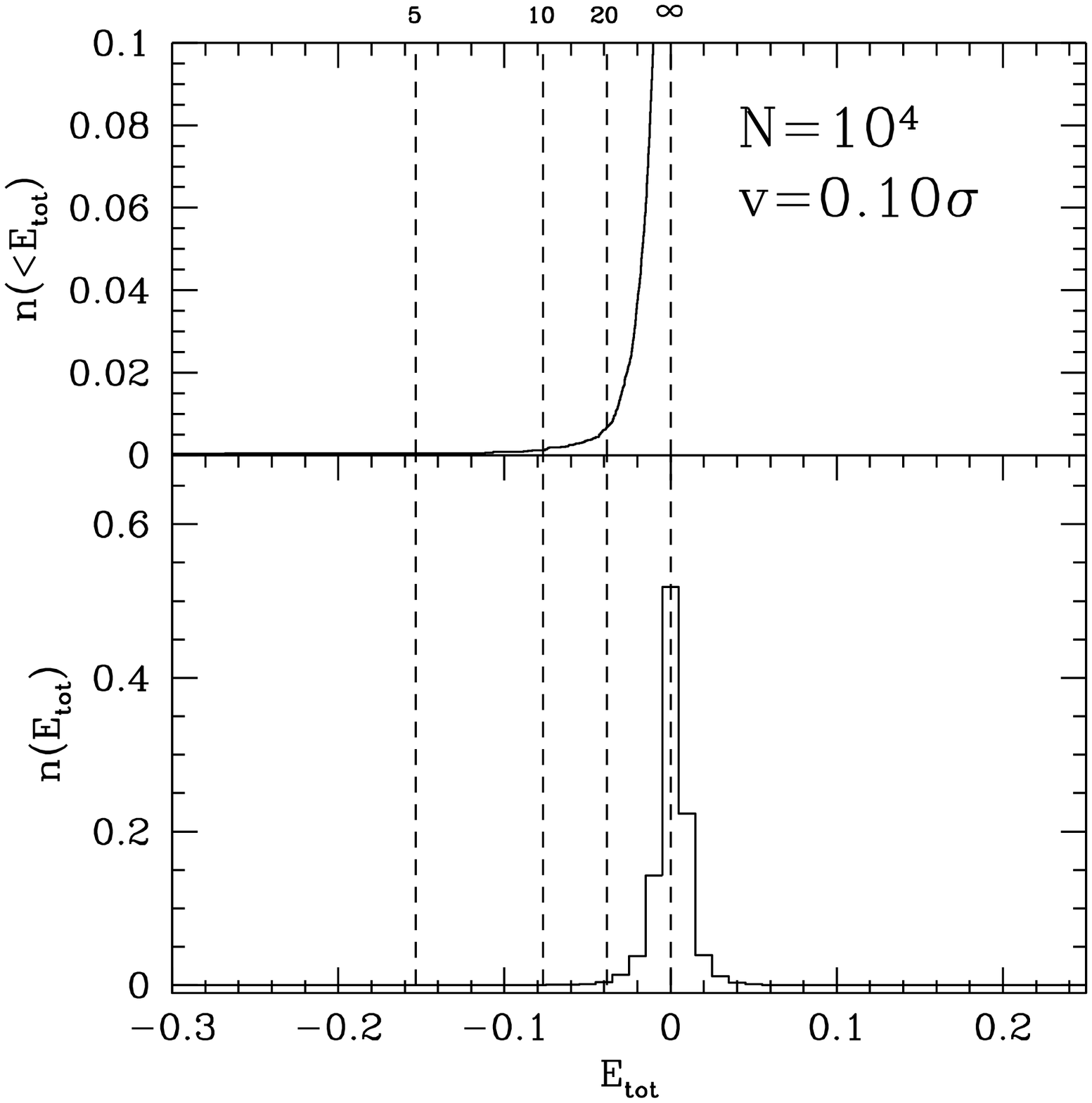,width=6.8cm}
\epsfig{figure=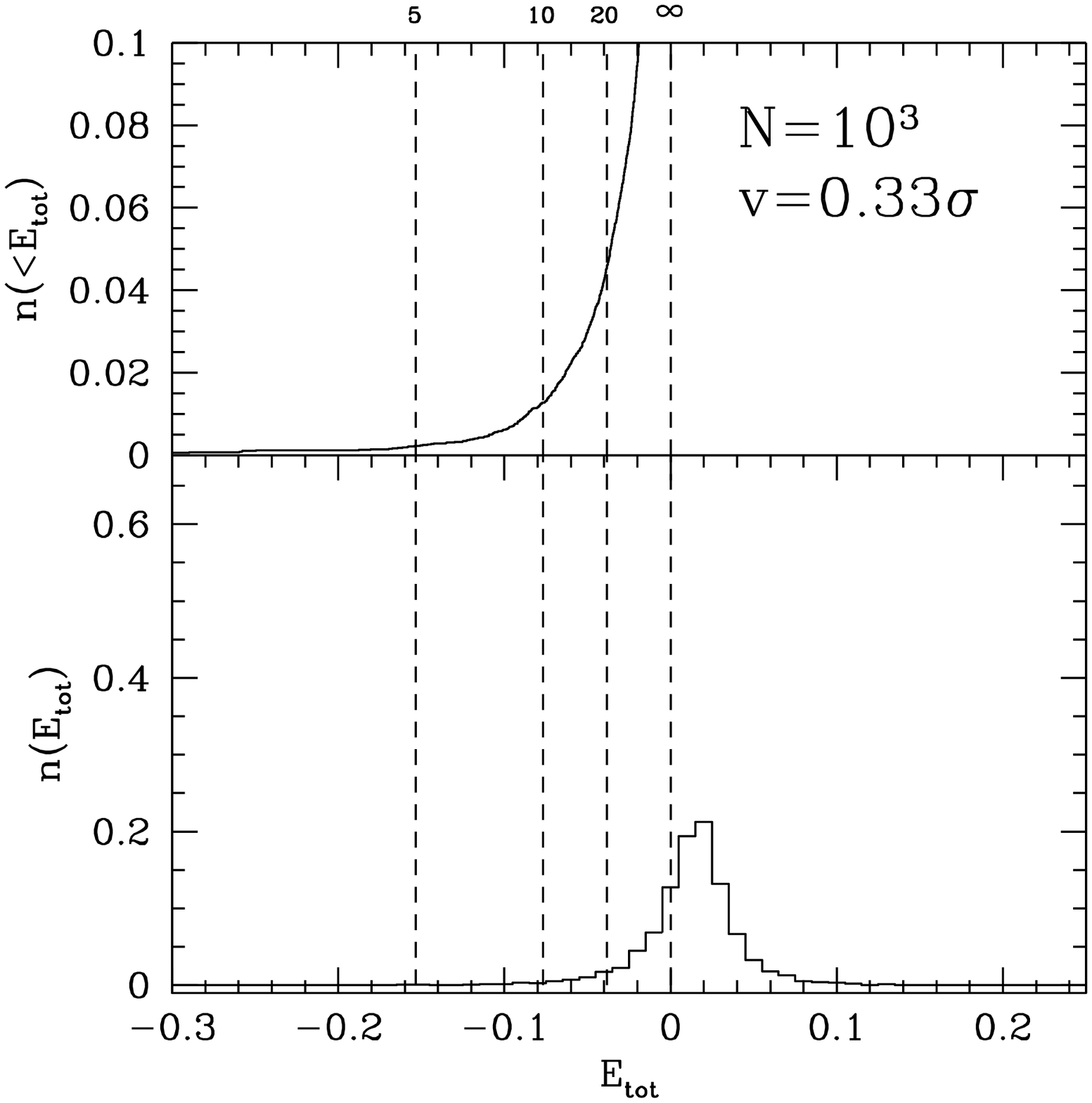,width=6.8cm}
\epsfig{figure=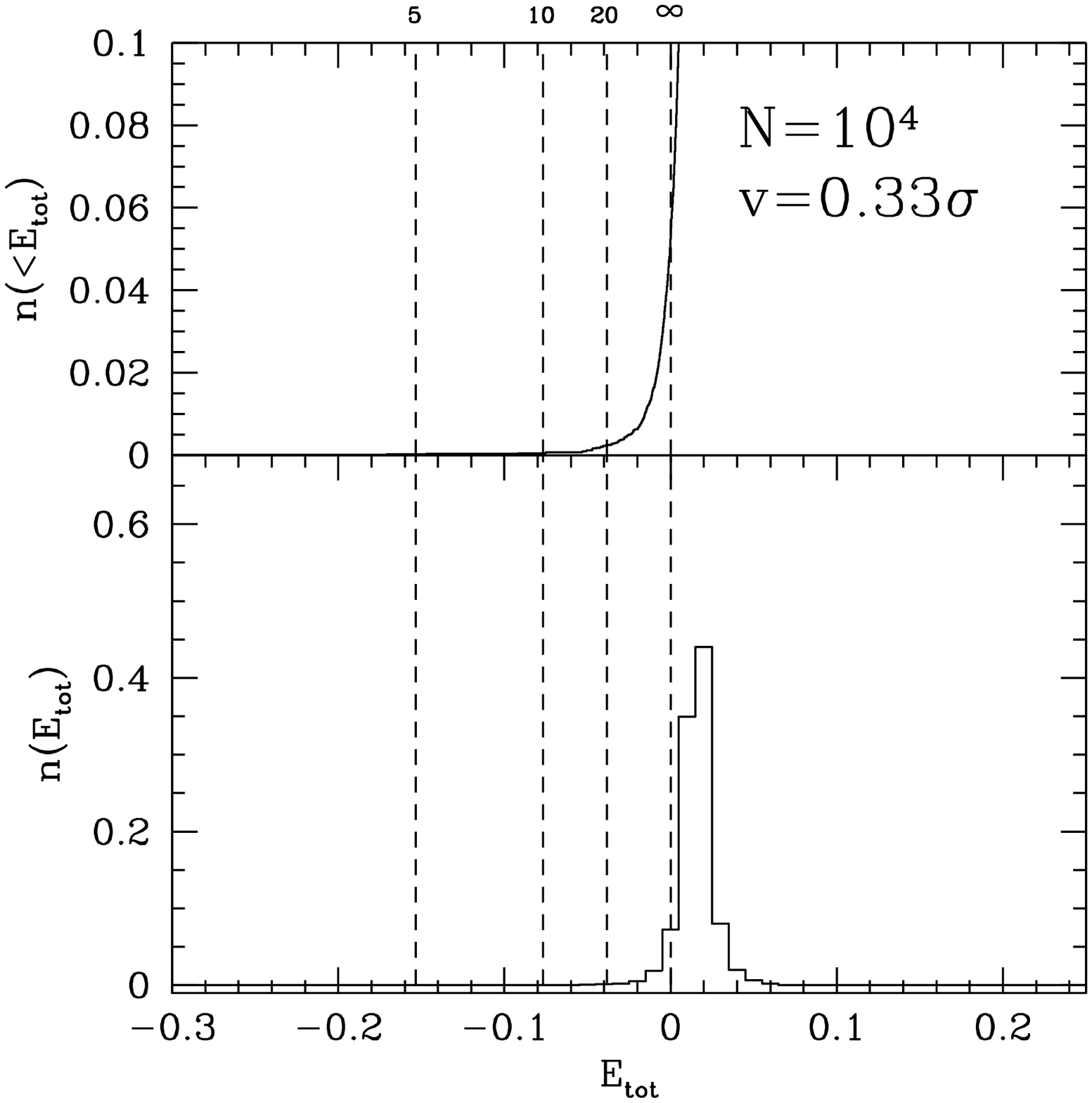,width=6.8cm}
\epsfig{figure=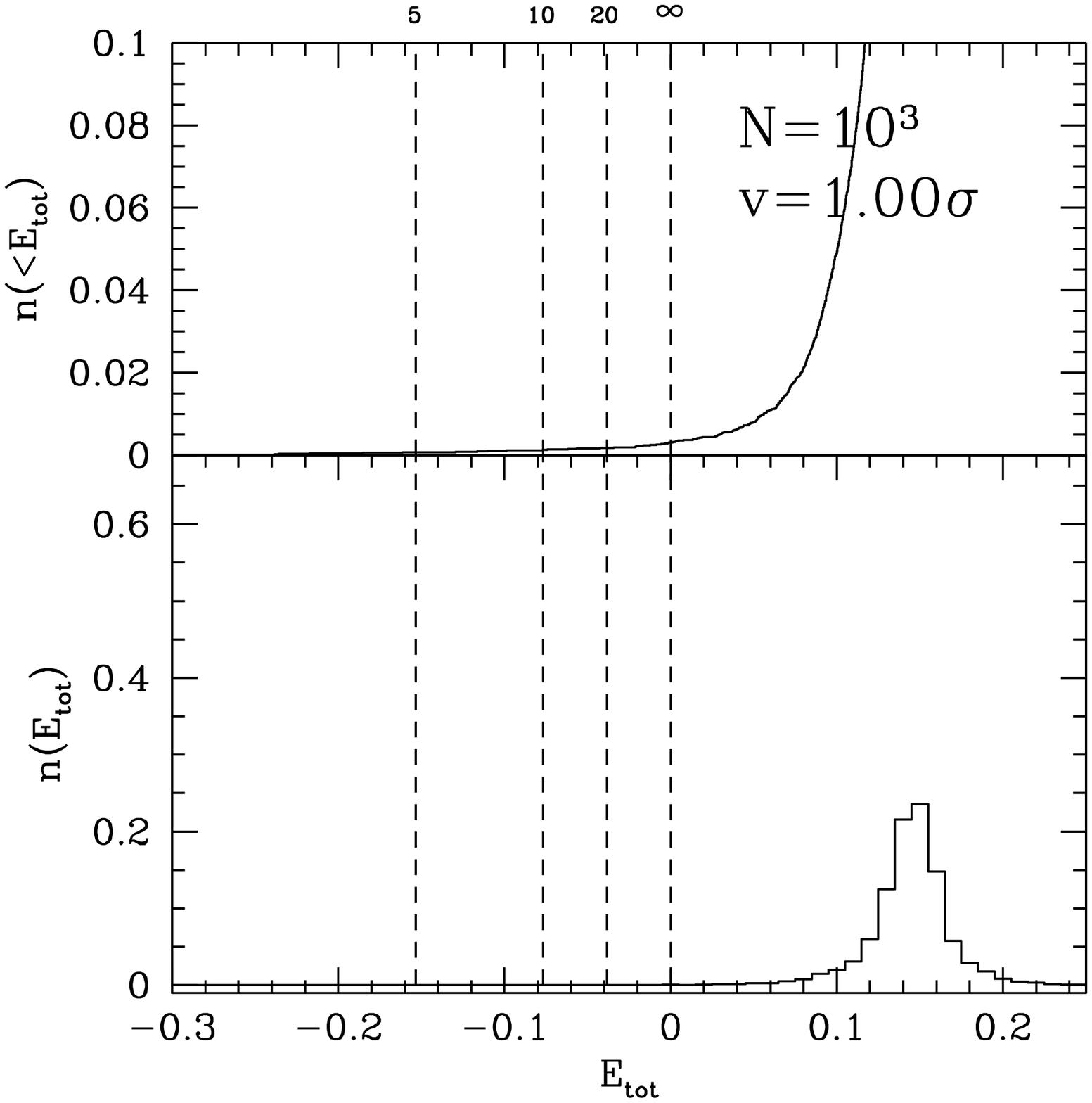,width=6.8cm}
\epsfig{figure=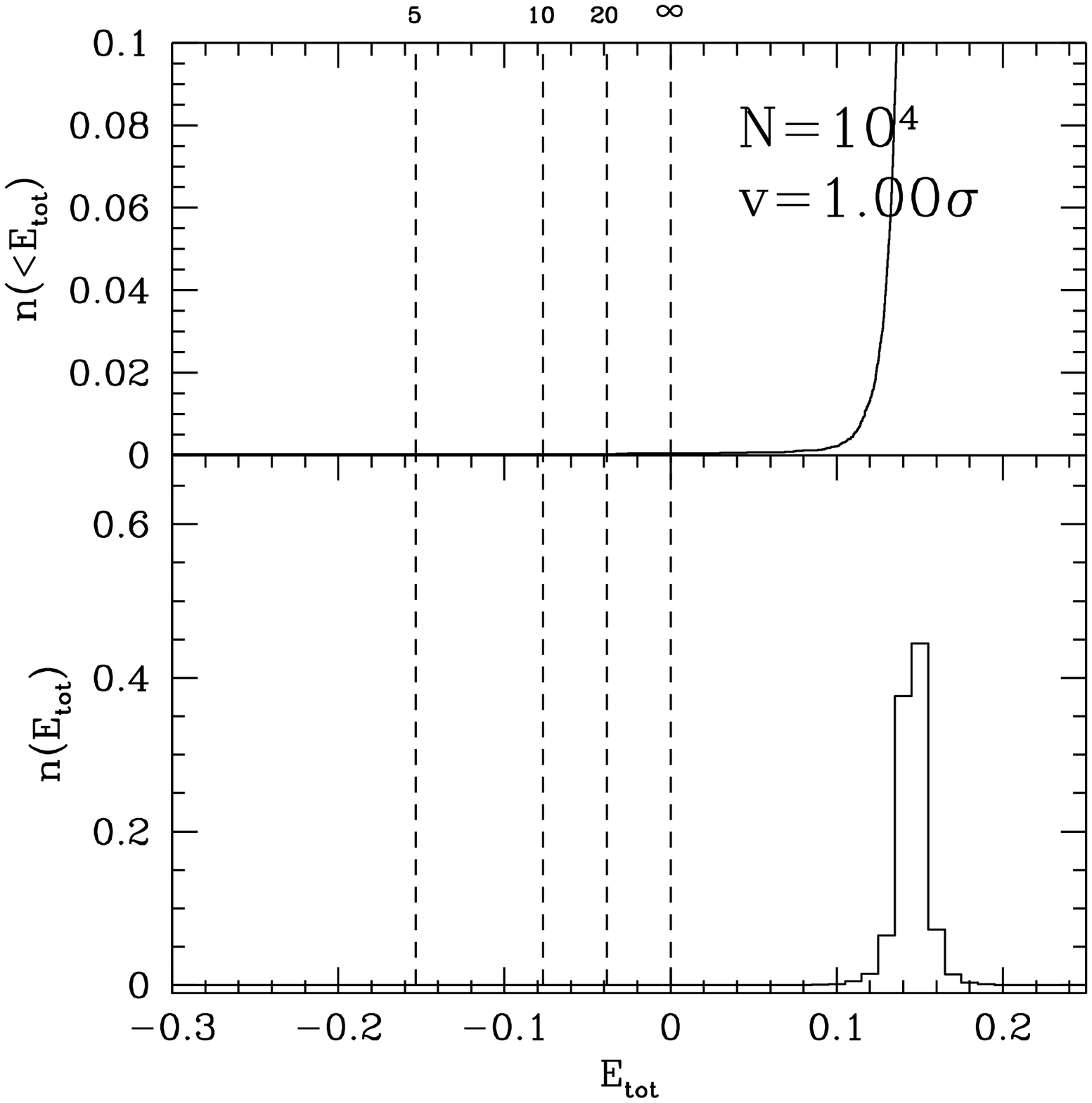,width=6.8cm}

\caption{Distribution of total energy E$_{tot}$ of field stars after interaction with the cluster stars. 
  {\bf Left panels:} Star clusters with $N$=10$^3$ stars. {\bf Right
    panels:} Star clusters with $N$=10$^4$ stars.  Bottom panels give
  the energy histogram, top panels give the cumulative stellar
  distribution. Indicated in each plot are the number of cluster stars
  and the field star initial velocities in units of the star cluster
  velocity dispersion $\sigma_{\rm cluster}$. The initial
energies corresponding to the three cases of v=1.0, 0.33 and 0.1 $\sigma$ are (in model units): 0.146, 0.0159, and 0.00146. They coincide with the peaks
of the respective energy distributions. The requirement for
  field star capture is $E_{tot}<\frac{-G M_c}{r_{\rm tid}}$, with
  $r_{\rm tid}$ being the tidal radius of the cluster in the
  gravitational field of the host galaxy. The vertical dashed lines
  indicate different assumed ratios of $\frac{r_{\rm tid}}{r_h}$,
  where $r_h$ is the cluster's half-mass radius.  }
\label{energy}
\end{center}
\end{figure*}

\begin{figure*}[h!]
\begin{center}
  \epsfig{figure=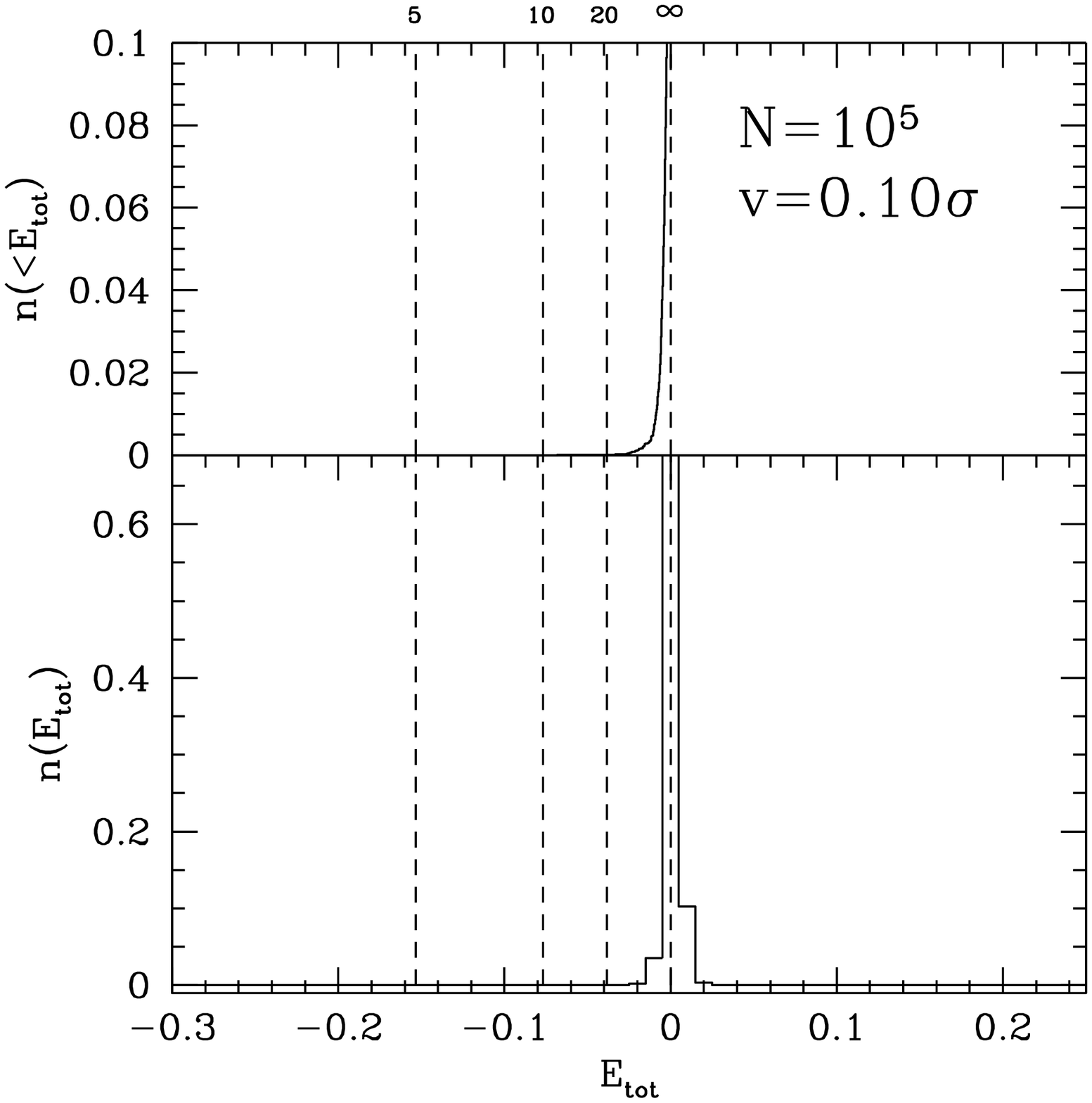,width=6.8cm}
\epsfig{figure=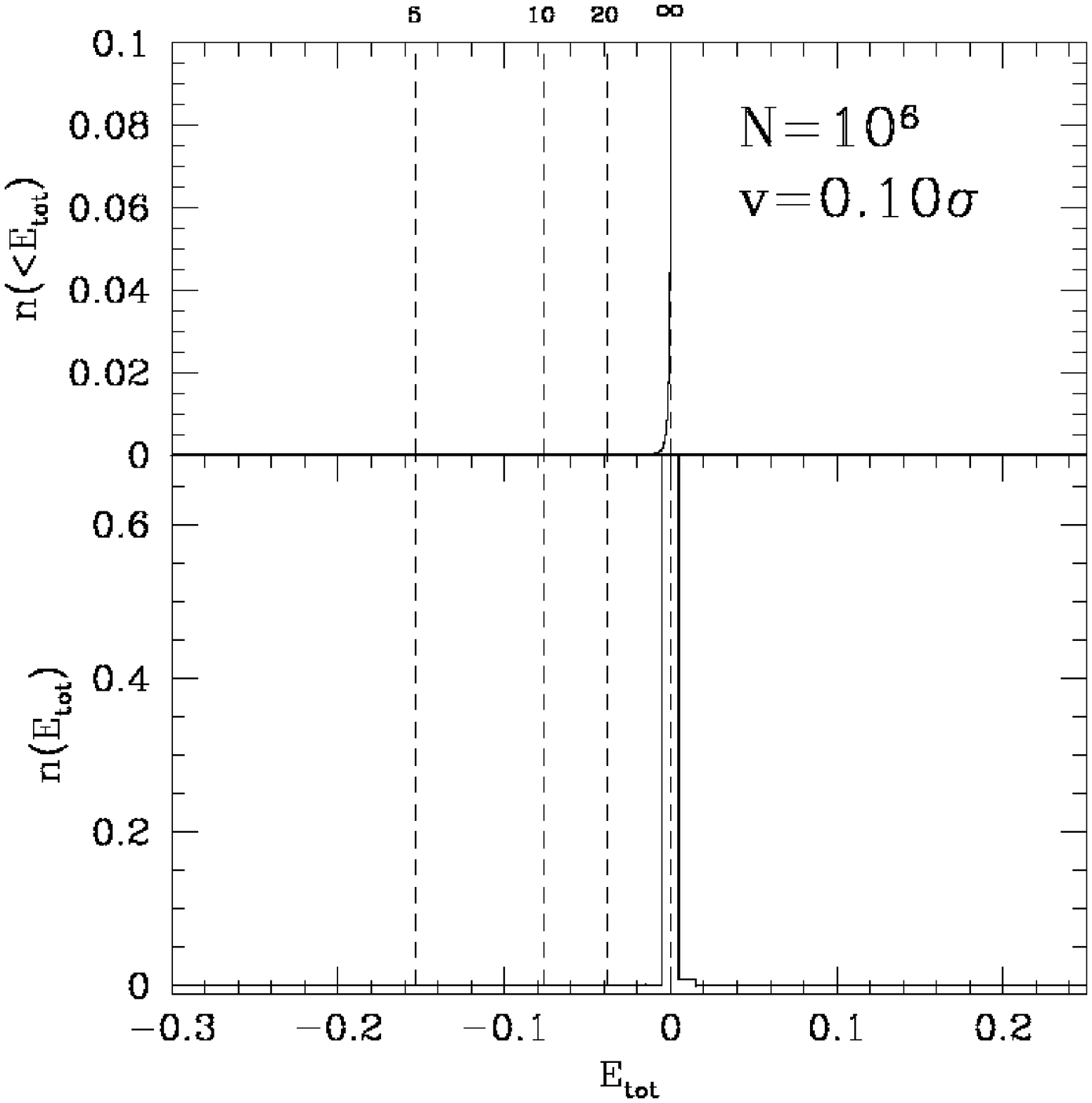,width=6.8cm}
\epsfig{figure=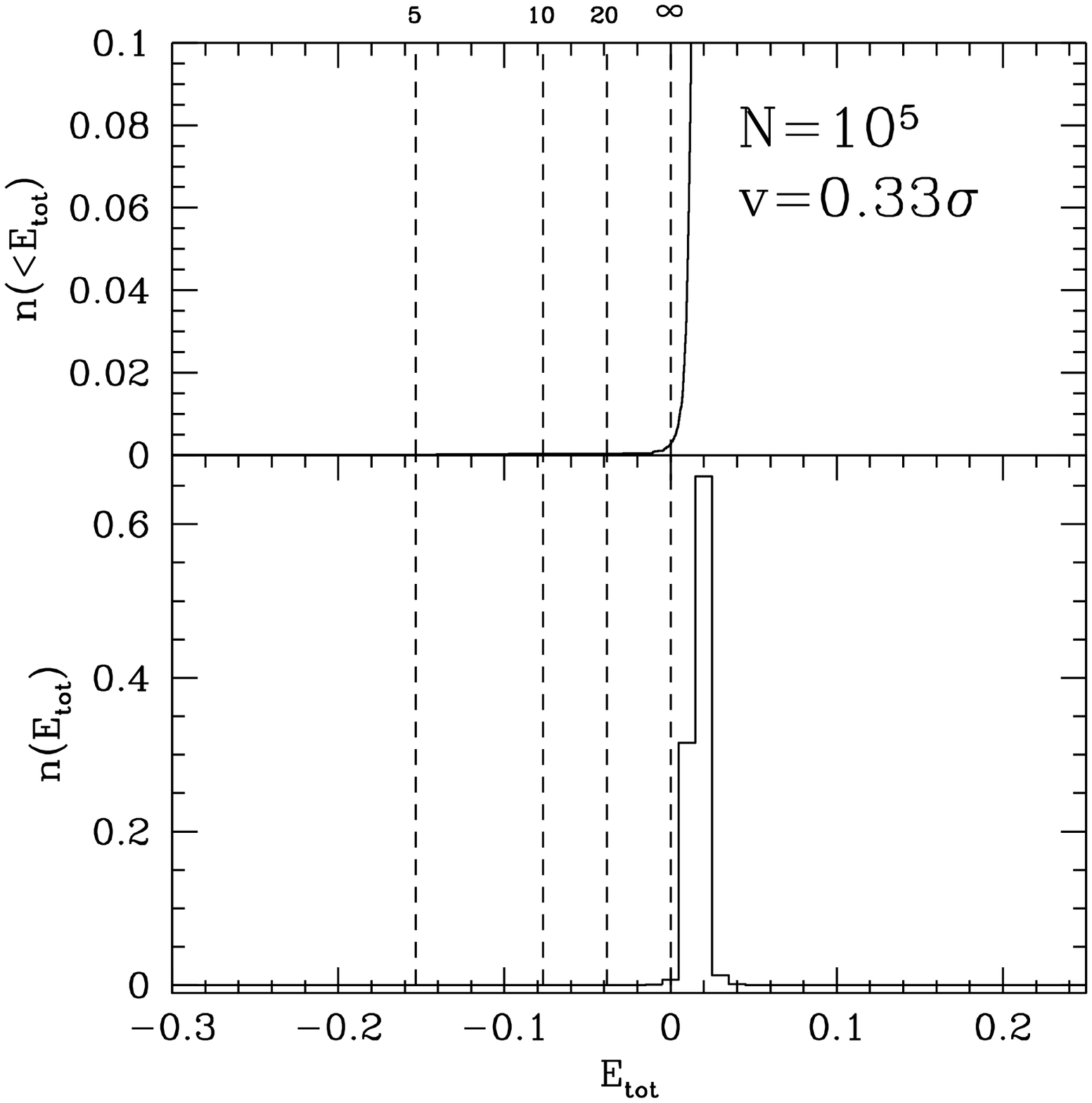,width=6.8cm}
\epsfig{figure=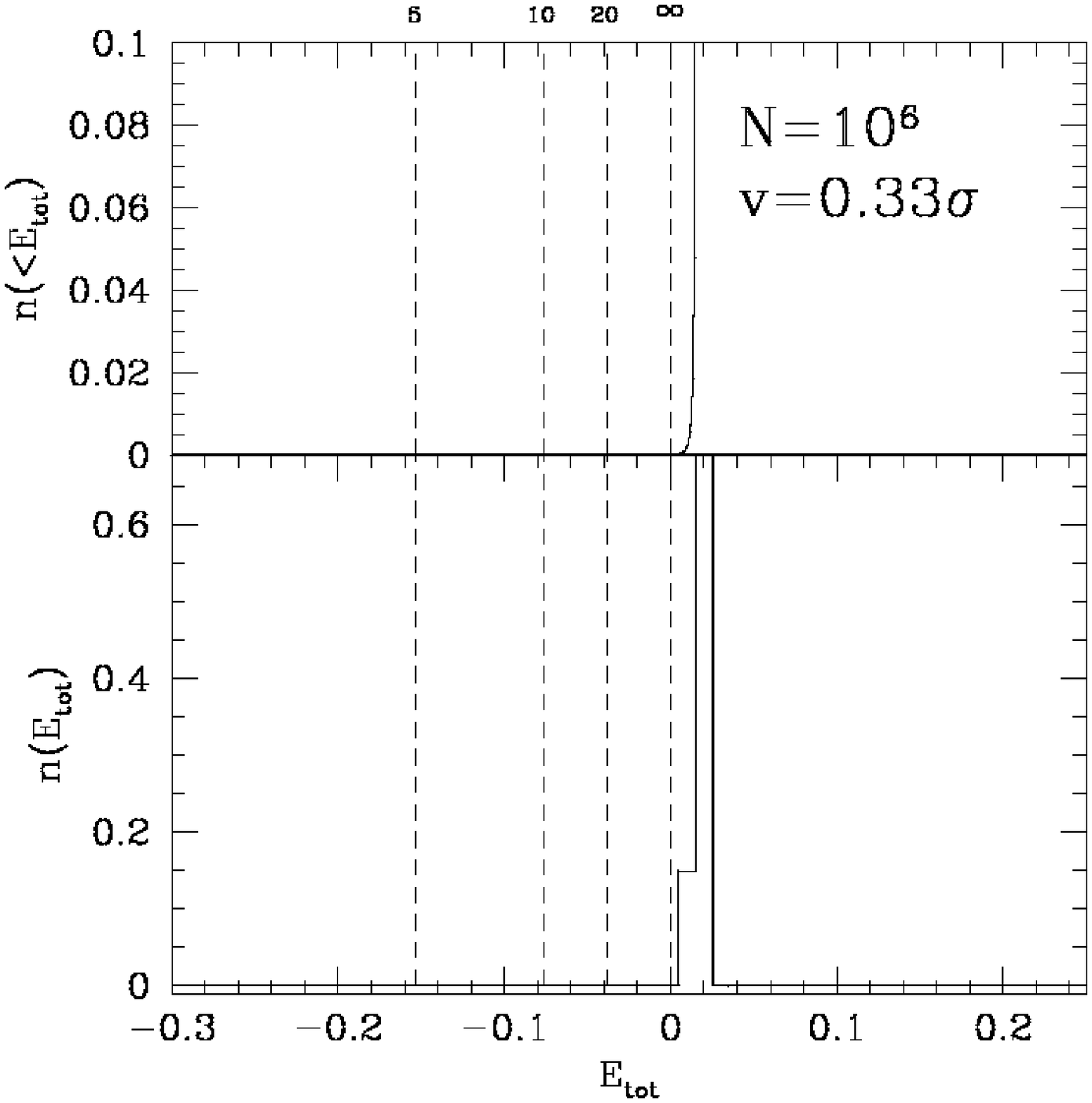,width=6.8cm}
\epsfig{figure=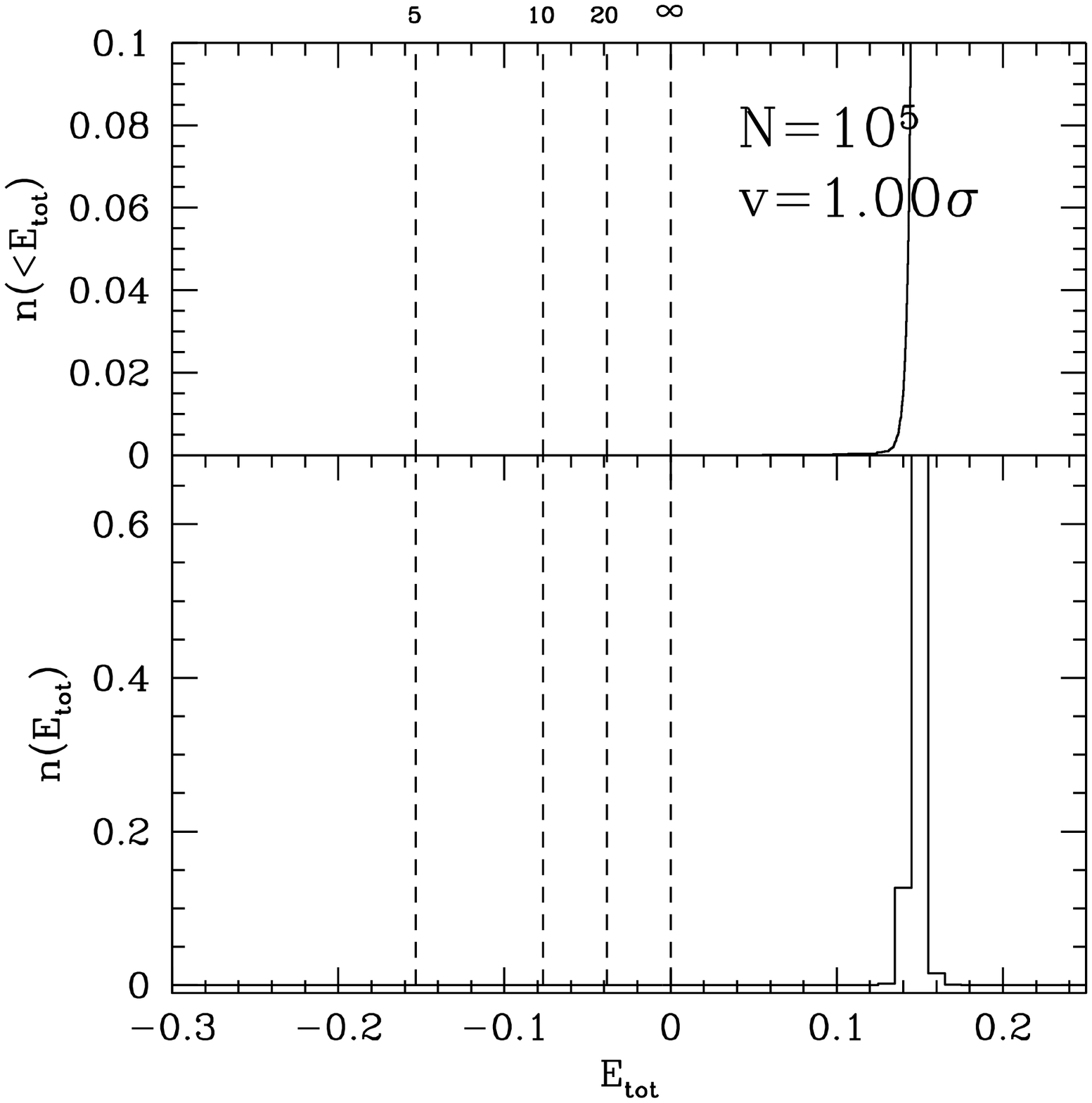,width=6.8cm}
\epsfig{figure=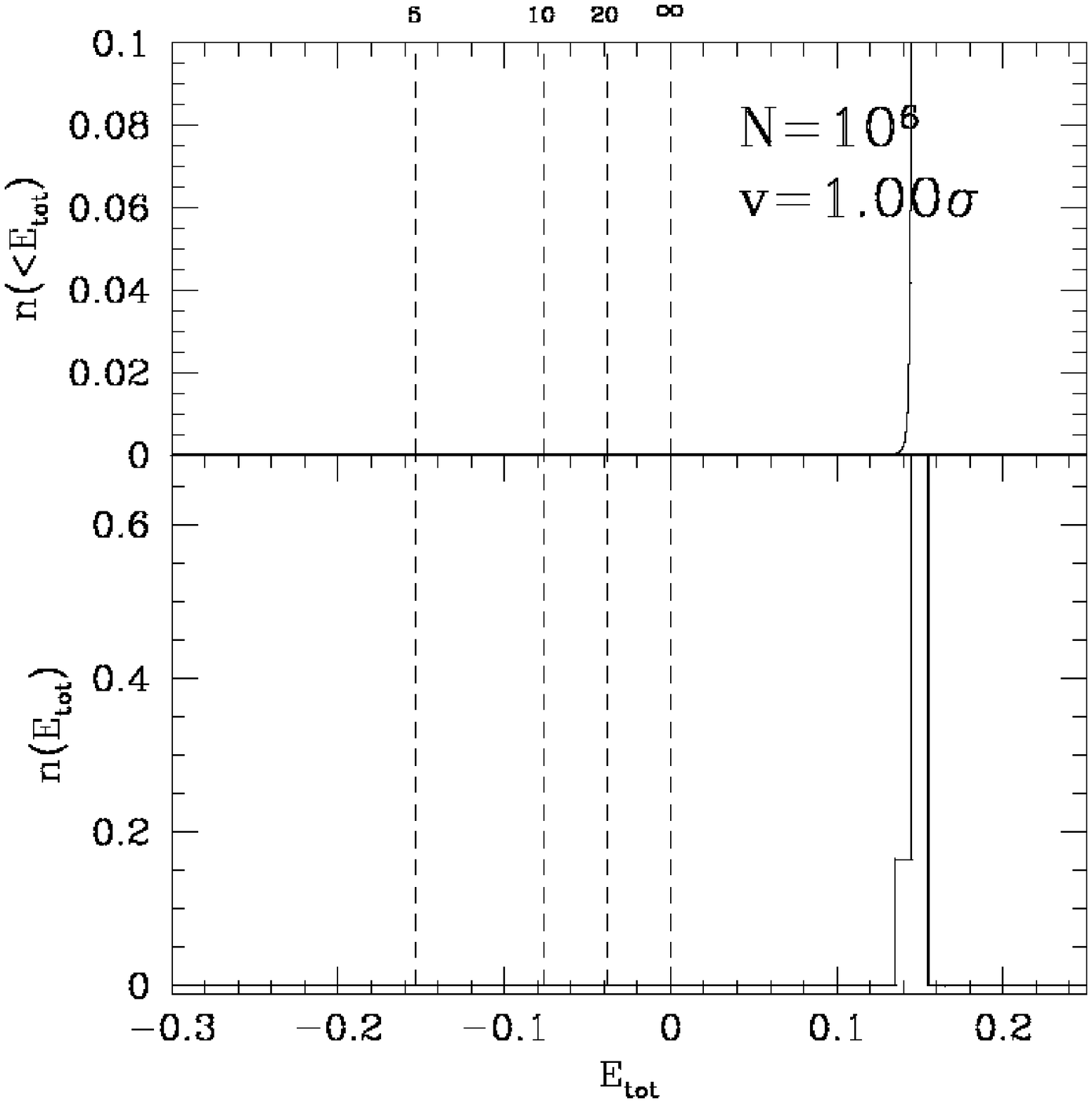,width=6.8cm}

\caption{Energy distributions of field stars after interaction like in Fig.~\ref{energy}, but here for clusters with $N$=10$^5$ (left) and 10$^6$ (right) stars. The curves for $N$=10$^6$ result from re-scaling the distribution of $N$=10$^5$ (see text). For high particle
  numbers, only a very small fraction of incoming field stars gets
  bound to the clusters.}

\label{energy5_6}
\end{center}
\end{figure*}

Figs.~\ref{energy} and \ref{energy5_6} show the energy distribution of
field stars after interaction with the cluster stars. The results for
clusters with $N$=10$^3$ to 10$^5$ stars are directly taken from the
corresponding N-body simulations. In Fig.~\ref{energy5_6} we also plot
results for $N=10^6$, where we have re-scaled the results from lower
$N$ (see below). This was done since calculations for large numbers of
cluster stars require exceedingly large calculation times.

In the units of the calculations shown in Figs.~\ref{energy}
and~\ref{energy5_6}, the mean kinetic energy of the cluster stars is
0.14582, which is equal to the total initial energy of field stars
when $v_{ini} = \sigma_{\rm cluster}$.

Due to the encounters between cluster and field stars, the energy
distribution of the field stars gets broadened.  If relaxation is
responsible for the broadening, one would expect that the dispersion
in energy of the field stars after the passage should be proportional
to
\begin{equation}
 \frac{\Delta E}{E} \sim \sqrt{\frac{t_{\rm Pass}}{t_{\rm rel}}}
\end{equation}
where $t_{\rm Pass}$ is the crossing time of the field stars through
the cluster and $t_{\rm rel}$ is the clusters relaxation time. Since
$t_{\rm rel} \sim \frac{N}{\ln{N}}$, the energy distribution should
become narrower for a larger number of cluster stars, which is indeed
observed in the runs.  For v$_{\rm ini}$=0.1$\sigma_{\rm cluster}$,
the width for $N$=10$^4$ is a factor of $\sqrt{9.4}$ smaller than for
$N$=10$^3$, and a factor of $\sqrt{9.1}$ larger than for $N$=10$^5$.
The corresponding factors for a scaling with the relaxation time are
$\sqrt{7.5}$ and $\sqrt{8.0}$, which is close to the observed scaling.
The reason for the remaining difference could be close enocunters
between cluster and field stars, which lead to large energy changes
and are not correctly described by relaxation. Since close encounters
are more important for low-mass clusters, the difference with the
observed scaling should become smaller for larger $N$, which is indeed
the case.

On the other hand, the width depends only very marginally on the
initial velocity: the width for v$_{\rm ini}$=0.1$\sigma_{\rm
  cluster}$ and $N$=10$^4$ is only 17\% broader than for v$_{\rm
  ini}$=1.0$\sigma_{\rm cluster}$ for $N$=10$^4$. Energy conservation
requires the velocity of a field star near the half-mass radius to be
given by $v_* = \sqrt{2 \sigma_{\rm cluster}^2 + v^2_{\rm ini}}$.
Since the crossing time is inversely proportional to $v_*$, we would
expect that the width is 22\% higher for v$_{\rm ini}$=0.1$\sigma_{\rm
  cluster}$, which is close to the above value.  Those well defined
dependencies allow us to make a robust extrapolation of the capture
probabilities from $N$=10$^5$ to $N$=10$^6$ to estimate
capture rates of massive globular clusters.

\begin{figure}[]
\begin{center}
  \epsfig{figure=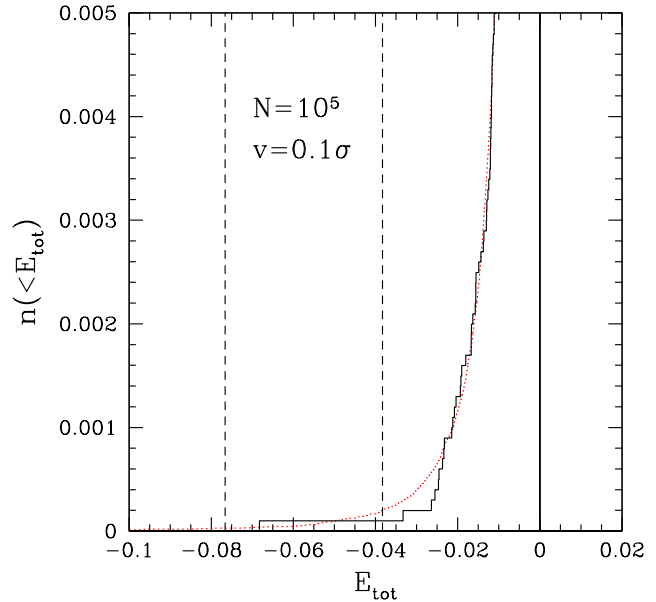,width=8.6cm}
\caption{
  Cumulative distribution of total energy E$_{tot}$ of field stars
  after interaction with the cluster stars for $M_c=0.5*10^5$M$_*$ and
  v=$0.1\sigma_{\rm cluster}$. Solid (black) histogram shows the
  result of N-body simulations. The dotted line is the cumulative
  version of a $t$-distribution chosen to fit the simulated
  distribution. This continuous function is used to extrapolate the
  capture probabilities for $N$=10$^6$ clusters (see text and
  Table~\ref{capture_prob}). The two vertical dashed lines indicate
  the energy required for capture for two different ratios between the
  cluster's tidal and half-mass radii. For the more negative energy,
  it is $\frac{r_{\rm tid}}{r_h}=10$. For the less negative energy, it
  is $\frac{r_{\rm tid}}{r_h}=20$. }
\label{compare}
\end{center}
\end{figure}

To obtain estimates for $N$=10$^6$ we re-scaled the energy
distributions for 10$^5$ by reducing its {\it width} by a factor of
$\sqrt{8.3}$. This is the reduction factor expected from $t_{\rm rel}
\sim \frac{N}{\ln{N}}$. From the comparisons above we would expect
this reduction factor to be a lower limit. Thus, the number of field
stars below a certain energy that are derived from this re-scaling
will be a slight over-estimate of the true number.  The {\it mean} of
the energy distributions for the various v$_{\rm ini}$ was adopted
identical to the values for $N$=10$^5$. This is possible because the
mean total energy of stars after interaction is identical to the
initial energy. 
The implicit assumption that tidal friction is negligible is discussed
in Sect.~\ref{discussion}.  In order to achieve a statistical
precision of the extrapolation to better than 10$^{-4}$ (i.e. the
inverse of the number of simulated stars), we fitted $t$-functions to
the energy distributions of the $N$=10$^5$ cluster and re-scaled this
continuous function to estimate the capture probability for a
$N$=10$^6$ cluster. For the fitting, we took special care to not underestimate
the wings of the energy distribution. See Fig.~\ref{compare} for a
comparison between the continuous fit and the discrete energy
distribution of field stars for the $N$=10$^5$ cluster and v$_{\rm
  ini}=0.1 \sigma_{\rm cluster}$.

In each plot of Figs.~\ref{energy} and \ref{energy5_6} we indicate the
potential energy at $r_{\rm tid}=$5, 10, and 20$\times r_h$.  These
values are typical tidal radii for open or globular clusters that have
a half-mass radius of a few pc and move on orbits which are a few kpc
away from the center of a Milky Way like galaxy.  The upper part of
each plot shows the cumulative energy distribution of the field stars
after the interaction. The fraction of captured field stars for a
given $r_{\rm tid}$ is determined by the intersection of the
cumulative distribution and the corresponding vertical dashed line.
The numerical values for $r_{\rm tid}$=5, 10 and 20 $\times r_h$ are given in
Table~\ref{capture_prob}. As can be seen, the capture probability is
extremely low especially for large initial velocities $\sim
\sigma_{\rm cluster}$.  However, even for small initial relative
velocities and large assumed tidal radii the probabilities are
negligible for most cases. Only for the very low mass case
($N$=10$^3$) of a small open cluster, the capture probabilites are a
few percent for field stars with initial velocities smaller than
$\sigma_{\rm cluster}$.


\begin{table}
\caption{Field star capture probabilities as function of field star initial velocity and number of cluster stars $N$. Probabilities refer to field stars with impact parameter $p<2r_h$. For $N$=10$^3$ to 10$^5$, the numbers are directly taken from N-body simulations of 10000 field star passages. For $N$=10$^6$, the probabilities are taken from a fit to the energy distribution of $N$=10$^5$ which was re-scaled to the expected width for $N$=10$^6$ (see text and Fig.~\ref{compare}).}
\label{capture_prob}
\begin{tabular}{ll|rrr}
$N$ &$r_{\rm tid}$& v$_{\rm ini}$=0.1$\sigma_{\rm cluster}$ &v$_{\rm ini}$=0.33$\sigma_{\rm cluster}$ &v$_{\rm ini}$=1.0$\sigma_{\rm cluster}$\\\hline\hline
10$^3$ & $5\times r_{\rm h}$ &4.7*10$^{-3}$ &2.2*10$^{-3}$ &7*10$^{-4}$\\
10$^3$ & $10\times r_{\rm h}$ &2.33*10$^{-2}$ &1.28*10$^{-2}$ &1.3*10$^{-3}$\\
10$^3$ & $20\times r_{\rm h}$ &8.49*10$^{-2}$ &4.55*10$^{-2}$& 1.7*10$^{-3}$\\\hline
10$^4$ & $5\times r_{\rm h}$ &4*10$^{-4}$ &2*10$^{-4}$ &2*10$^{-4}$\\
10$^4$ & $10\times r_{\rm h}$ &1.1*10$^{-3}$ &5*10$^{-4}$ &2*10$^{-4}$\\
10$^4$ & $20\times r_{\rm h}$ &6.7*10$^{-3}$ &2.4*10$^{-3}$ & 2*10$^{-4}$\\\hline
10$^5$ & $5\times r_{\rm h}$ & $<5*10^{-5}$ & $1*10^{-4}$  & $<5*10^{-5}$\\
10$^5$ & $10\times r_{\rm h}$ & $<5*10^{-5}$ & $3*10^{-4}$  & $<5*10^{-5}$\\
10$^5$ & $20\times r_{\rm h}$ & 1*$10^{-4}$ & $3*10^{-4}$ & $<5*10^{-5}$\\\hline
10$^6$ & $5\times r_{\rm h}$ & $3*10^{-7}$ & $3*10^{-7}$  & $<10^{-7}$\\
10$^6$ & $10\times r_{\rm h}$ & $1.9*10^{-6}$& $1.4*10^{-6}$& $<10^{-7}$\\
10$^6$ & $20\times r_{\rm h}$ & $1.2*10^{-5}$  & $5*10^{-6}$ & $1*10^{-7}$\\\hline\hline
\end{tabular}
\end{table}

\subsection{Capture rates}
To transform the capture probabilities into actual capture rates, we
need to know the number of 'fly-by' field stars. These are those stars
that over 10 Gyr approach the cluster within 2*$r_{\rm h}$ and within
the various $v_{\rm ini}$ ranges. It is important to re-iterate that
the initial velocities $v_{\rm ini}$ in Table~\ref{capture_prob} are
expressed in units of star cluster velocity dispersion $\sigma_{\rm cluster}$.
Depending on which mass is assumed for a single star, these ranges
correspond to different ranges in km/s. In the following we adopt as
mass of a single star 0.5 solar masses, which is the typical average
mass of stars in clusters of the investigated range. The mass range
of the investigated cases hence ranges from 0.5 * 10$^3$ to 0.5 * 10$^6$
M$_{\sun}$. This covers the regime from open clusters up to globular clusters
one magnitude more massive than the mass-function turn-over (2*10$^5$ M$_{\sun}$,
e.g. Jord\'{a}n {\it et al.}~\cite{Jordan07}).

For calculating the fly-by rates, we assume a typical globular cluster
half-mass radius of $r_h=3$ pc (Jord\'{a}n et
al.~\cite{Jordan05}~and~\cite{Jordan07}). Furthermore, we adopt a
field star density of 0.1 $L_{\sun} pc^{-3}$, which is comparable to
typical values in giant elliptical galaxies within $r_{\rm eff}$
(Romanowsky {\it et al.}~\cite{Romano01}), and also similar to the value in the solar
neighbourhood. Assuming a typical M/L ratio of 2.5, this corresponds to a mass
density of 0.25 $M_{\sun} pc^{-3}$. 

The resulting fly-by rates are shown in
Table~\ref{fieldstar_rate}.  The calculations are done for masses
between 0.5*10$^3$ and 0.5*10$^6$ $M_{\sun}$, and for a range of
relative velocity dispersions between field stars and cluster. The
highest velocity dispersion (485 km/s) represents the case of a GC
orbiting a giant elliptical galaxy like M87 (see Mieske et
al.~\cite{Mieske06}).  The lowest velocity dispersion (15 km/s)
represents co-rotation of an open or globular cluster in a dynamically
cold disk.  The number of fly-by stars is given as fraction of the
number of stars in the cluster. For high field star velocity
  dispersions and low cluster masses, not a single field star
  with $v<\sigma_{\rm cluster}$ passes the cluster over a Hubble time.

The capture probabilities (assuming 0.5 solar mass stars) convolved with
the number of ``fly-bys'' within 10 Gyrs yields the mass fraction of
captured field stars. These numbers are shown in
Table~\ref{fieldstar_conv} for the case of $r_{\rm tid}=$20$\times
r_h$. For most cases, the number of captured field stars is lower than
1 single star over a Hubble time. Only for the more massive examples 
0.5*10$^4$ to 0.5*10$^6$ $M_{\sun}$ and the cold disk case, up to a few dozen
stars will be captured. The mass fraction of captured stars is $\le 2*10^{-4}$
 for all considered cases.
\vspace{0.1cm}

This shows that field star capture over a Hubble-time will not change
the integrated photometric parameters of a star cluster.


\begin{table*}
\caption{Fraction of field stars over cluster stars f($\sigma_{\rm field}$) that approach a star cluster within 10 Gyr and within
2*$r_{\rm h}$, within adjacent ranges around $v_{\rm ini}$ indicated in column 3. $\frac{v_{\rm ini}}{\sigma_{\rm cluster}}=0.1$ corresponds to the range [0 : 0.18]. $\frac{v_{\rm ini}}{\sigma_{\rm cluster}}=0.33$ corresponds to [0.18 : 0.57]. $\frac{v_{\rm ini}}{\sigma_{\rm cluster}}=1$ corresponds to [0.57 : 1]. 
The star cluster is assumed to have $r_{\rm h}=3pc$, consist of 0.5 solar mass stars, and be embedded 
within a field star density of 0.1 $L_{\sun} pc^{-3}$. The approach fractions (or ``fly-by'' rates) are calculated for different 
assumed velocity dispersions of the field stars $\sigma_{\rm field}$.}

\label{fieldstar_rate}
\begin{center}
\begin{tabular}{lll|rrrr}
Mass [M$_{\sun}$]&$\sigma_{\rm cluster}$ [km/s] &$\frac{v_{\rm ini}}{\sigma_{\rm cluster}}$&  f($\sigma_{\rm field}=485$ km s$^{-1}$)& f($\sigma_{\rm field}=200$ km s$^{-1}$)& f($\sigma_{\rm field}=50$ km s$^{-1}$)&  f($\sigma_{\rm field}=15$ km s$^{-1}$)\\\hline\hline
0.5*10$^3$ & 0.4 & 0.1&0        &0      &0      &1*10$^{-3}$     \\
0.5*10$^3$ & 0.4 & 0.33&0        &0      &0      &5*10$^{-3}$  \\
0.5*10$^3$ & 0.4 & 1  &0        &0      &0 &1.2*10$^{-2}$ \\\hline

0.5*10$^4$ & 1.3 & 0.1&0        &0      & 1.5*10$^{-4}$ & 5*10$^{-3}$\\
0.5*10$^4$ & 1.3 & 0.33&0        &0      &1.4*10$^{-3}$ &5*10$^{-2}$\\
0.5*10$^4$ & 1.3 & 1  &0        & 0&3.6*10$^{-3}$ & 0.13 \\\hline

0.5*10$^5$ & 4.2   & 0.1  &   0    &2*10$^{-5}$ &1.5*10$^{-3}$  &  6*10$^{-2}$  \\
0.5*10$^5$ & 4.2   & 0.33  &1.6*10$^{-5}$ &2*10$^{-4}$ &1.4*10$^{-2}$  &   0.52   \\
0.5*10$^5$ & 4.2   & 1  &4*10$^{-5}$ &6*10$^{-4}$ &3.8*10$^{-2}$  &   1.4   \\\hline

0.5*10$^6$ & 13.4  & 0.1  &1.7*10$^{-5}$  &2*10$^{-4}$ &1.6*10$^{-2}$ &  0.58 \\
0.5*10$^6$ & 13.4  & 0.33  &1.6*10$^{-4}$  &2*10$^{-3}$  &0.14  &  5   \\
0.5*10$^6$ & 13.4  & 1    &4*10$^{-4}$  &6*10$^{-3}$  &0.37   &  11   \\\hline\hline
\end{tabular}
\end{center}\end{table*}

\begin{table*}
\caption{Number ratio $n$ of captured stars after 10 Gyrs over total number of star cluster stars for initial velocities $v_{\rm ini}\le 1\sigma_{\rm cluster}$ and a range of assumed field star velocity dispersions $\sigma_{\rm field}$. The given values of $\sigma_{\rm cluster}$ are for the case that the simulated stars have 0.5 solar masses. For the tidal radius we assume $r_{\rm tid}=20*r_h$. The table is created by multiplying the field star fly-by rates from Table~\ref{fieldstar_rate} with the capture probabilities from Table~\ref{capture_prob}, and adding up the figures for ratios $\frac{v_{\rm ini}}{\sigma_{\rm cluster}}=$0.1, 0.33, and 1.0. A ratio of $n=0$ statistically corresponds to
$<$0.5 captured stars.}
\label{fieldstar_conv}
\begin{center}
\begin{tabular}{ll|rrrr}
Mass [M$_{\sun}$] &$(\sigma_{\rm cluster}$ [km/s]) & $n(\sigma_{\rm field}=485$ km s$^{-1}$)& $n(\sigma_{\rm field}=200$ km s$^{-1}$)& $n(\sigma_{\rm field}=50$ km s$^{-1}$)&  $n(\sigma_{\rm field}=15$ km s$^{-1}$)\\\hline\hline
0.5*10$^3$ & 0.4 &0        &0      &0 &0  \\
0.5*10$^4$ & 1.3 &0        &0 & 0& $2*10^{-4}$  \\
0.5*10$^5$ & 4.2   &0 &0 & 0&   $2.4*10^{-4}$  \\
0.5*10$^6$ & 13.4  &0 &0  & 0   &  $3.3*10^{-5}$   \\\hline\hline
\end{tabular}
\end{center}\end{table*}

\section{Discussion and conclusions}
\label{discussion}
In Bica {\it et al.}~(\cite{Bica97}), the capture of field stars by a globular
cluster orbiting the Milky Way bulge was calculated both analytically
and by means of simulations. Those authors find that over 1 Gyr, the
number of captured field stars by a star cluster of 0.5*10$^5$ $M_{\sun}$
is of the order of a few to 10\% that of the cluster stars. This relatively
high fraction of captured stars is in harsh contrast to our findings.
Why is this?

The reason lies in a fundamentally different approach towards
estimating the capture rate. In the simulations by Bica {\it et al.}, the
star cluster is placed ad hoc into a cloud of randomly moving field
stars. Field stars which then happen to be located within the radius
of the cluster and with relative velocity below escape velocity will
be captured. We note that also in M06 we
adopted this approach to analytically estimate the number of captured
field stars.

This approach neglects the fact that field stars feel the
gravitational potential of the star cluster already before they
``enter'' the cluster. By definition, even a field star with a zero
initial velocity at infinity will get attracted by the star cluster
such that its kinetic energy upon reaching the cluster center is
exactly equal to the necessary escape energy. This means that in absence
of energy exchange with star cluster stars and in absence of a time
variance of the cluster potential, {\it no field star will be
captured}. It is the two-body encounters that are required for any field
star to obtain a negative energy. The very small number of encounters
with a large energy transfer results in such low capture rates as presented
in the present paper. 

We compare our capture probabilities with a more recent study by
  Mints, Glaschke \& Spurzem~(\cite{Mints07}). Those authors
  investigate field star scattering and capture by open clusters that
  have 200, 500, and 2000 stars. They apply analytical estimates as
  well as Monte Carlo and N-body simulations, which agree well with
  each other. There are two differences between their approach and
  ours. First, Mints {\it et al.}  consider head-on collisions, that
  is impact parameters $p=0$. Our estimates for the capture
  probability cover the realistic range of $p<2*r_h$ (see
  Fig.~\ref{radius}), and should thus yield at most equal or lower
  capture probabilities.  Second, Mints {\it et al.} use $E<0$ as
  capture criterion. We impose $E<\frac{-G M_c}{r_{\rm tid}}$, taking
  into account the finite gravitational sphere of influence of a star
  cluster. Also this difference decreases the capture probabilities
  derived by us.
  
For comparison with their results, we have to consider the lowest
number case of N=1000 in our data, and also use $E<0$ as capture
condition.  While the Mints {\it et al.} cases of N=500 and N=2000 are
equally distant to N=1000 in logarithmic space, a comparison with our
data is more realistic for the N=2000 case. This is because the
capture probabilities of Mints {\it et al.} will be higher than ours due to
the assumption of head-on collisions.  Since the capture probabilities
decrease with increasing star number, the Mints {\it et al.}  case for
N=2000 should thus correspond best to our case of N=1000.  For the
three values $v_{\rm ini}=0.1, 0.33, 1.0$ $\sigma_{\rm cluster}$, we
find probabilities for $E<0$ in our data of 0.498, 0.244, and 0.003,
respectively. The respective probabilities from the Mints {\it et al.}
study are 0.48-0.5, 0.35, and 0.006 (the Monte Carlo simulations from Fig. 5
of their paper). The values agree well, lending independent
support to the conclusions drawn in the present paper.  

\begin{figure}[h]
\begin{center}
  \epsfig{figure=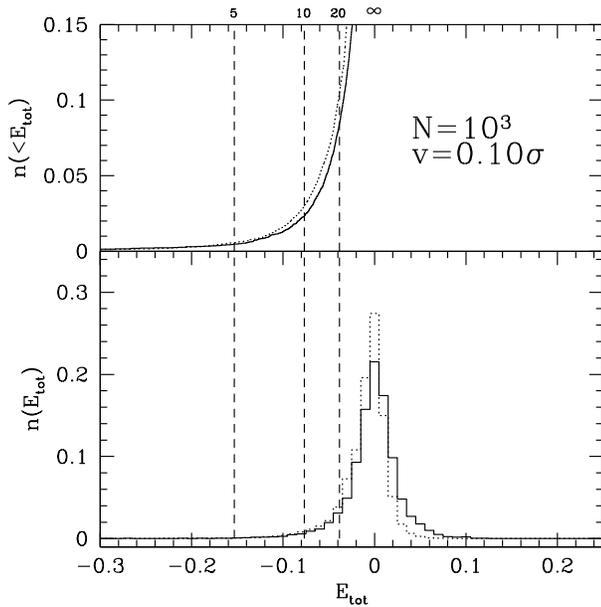,width=8.6cm}
\caption{Comparison of field star energy distribution after cluster interaction for the default case of equal mass for field star and cluster star(solid lines), and 5 times more massive field stars (dotted lines). The simulations are for a $N$=10$^3$ cluster and v$_{\rm ini}=0.1\sigma_{\rm cluster}$, cf. upper left panel of Fig.~\ref{energy}.}
\label{energy_1e3_compM5}
\end{center}
\end{figure}

Baranov~(\cite{Barano75}) argue that tidal friction causes field stars
to loose considerable amounts of energy during passage through a
cluster, and hence become captured. Tidal friction would make itself
note by a net energy loss of field stars after passage through the
cluster. For the simulations presented in Fig.~\ref{energy}, there is
no notable effect of tidal friction. This is because the simulated
field stars and cluster stars have the same mass. We have tested the
influence of tidal friction by performing one simulation for field
stars with 5 times the mass of a cluster star, for the case of a
$N$=10$^3$ cluster and v$_{\rm ini}=0.1 \sigma_{\rm cluster}$. In
Fig.~\ref{energy_1e3_compM5}, the resulting energy distribution of
field stars is compared with the same setup for equal mass field
stars. The effect of tidal friction is notable: the energy
distribution of the more massive field stars is skewed towards more
negative values. However, the effect is relatively small. The amount
of captured stars increases by only 20\% compared to the equal mass
case. We can therefore state that the effect of tidal friction is not
important in our simulations.

Investigating possible reasons for the multiple stellar
  populations found in several Milky Way globular clusters, Fellhauer
  et al.~(\cite{Fellha06}) showed that during the time of cluster
  formation -- i.e. when the cluster potential gets deeper with
    time due to the contraction of a gas cloud -- a significant
  amount of field stars may be trapped. However, this early trapping
cannot explain the colour-magnitude trend in GCs of early-type
elliptical galaxies.  This is because the field stars captured at
cluster formation would likely be more metal-poor than the newly
formed stars in the GC.  Capture at later times, when the field star
populations have already become metal-enriched in the course of
galaxy-mergers, is required to explain the 'blue tilt'.

\vspace{0.4cm}

\noindent We conclude that field star capture over a Hubble-time will not 
change the integrated photometric parameters of a star cluster,
provided that the gravitational potential of the cluster changes only
slowly compared to the cluster crossing time. Field star capture 
is not a probable mechanism for creating the colour-magnitude
trend of old metal-poor globular clusters. 


\acknowledgements
We thank the referee for his comments and A. Mints for a useful discussion.

\end{document}